\documentclass[showpacs,aps,prb,twocolumn]{revtex4}
\usepackage{amssymb,amsmath,bm,graphicx,color}

\begin{document}

\title{Role of dimensionality in spontaneous magnon decay: easy-plane
ferromagnet}
\author{V. A. Stephanovich}
\affiliation{Institute of Physics, Opole University, ul. Oleska 48, Opole, 45-052, Poland}
\author{M. E. Zhitomirsky}
\affiliation{Service de Physique Statistique, Magn\'etisme et Supraconductivit\'e,
UMR-E9001, CEA-INAC/UJF, 17 rue des Martyrs, 38054 Grenoble cedex 9, France}

\date{\today}

\begin{abstract}
We calculate magnon lifetime in an easy-plane ferromagnet on a tetragonal
lattice in transverse magnetic field. At zero temperature  magnons are unstable
with respect to spontaneous decay into two other magnons. Varying ratio of intrachain
to interchain exchanges in this model we consider the effect of dimensionality on
spontaneous magnon decay.
The strongest magnon damping is found in the quasi-one-dimensional case for momenta
near the Brillouin zone boundary. The sign of a weak interchain coupling has a
little effect on the magnon decay rate. The obtained theoretical results  suggest possibility
of experimental observation of spontaneous magnon decay  in a quasi-one-dimensional
ferromagnet CsNiF$_3$. We also find an interesting enhancement of the magnon decay rate
for a three-dimensional ferromagnet. The effect is present only for the nearest-neighbor model
and is related to effective dimensionality reduction in the two-magnon continuum.
\end{abstract}
\maketitle

\section{Introduction}

Magnons are commonly viewed as bosonic quasiparticles with integer spin $S^z=1$.
This is certainly true for isotropic ferromagnets, which were originally treated by
Felix Bloch in his seminal paper. \cite{Bloch30} In the isotropic case the total
spin (magnetization) is conserved and magnon interaction in an isotropic ferromagnet
amounts to simple particle-particle scattering or four-magnon processes.
\cite{Holstein40,Oguchi60} In the presence of magnetic anisotropy, {\it e.g.},
dipolar or single-ion, the total spin is no longer conserved and definite
spin of a magnon ceases to exist as well. As a result, additional three-particle
interaction terms appear in the magnon Hamiltonian. \cite{Holstein40,Akhiezer46} Spin waves
in antiferromagnets have no definite value of $S^z$ even in the isotropic case since
the quantum ground state is now a superposition of states with different total spins.
\cite{Anderson} Still three-magnon processes appear only in noncollinear antiferromagnetic
structures with completely broken spin-rotational symmetry, \cite{Ohyama93,Zhitomirsky99}
whereas magnon-magnon interactions in collinear antiferromagnets are represented by particle
non-conserving four-magnon processes. \cite{Oguchi60,Harris71}

A special role of three-magnon dipolar processes for spin relaxation in ferromagnets
was recognized already in the early works. \cite{Akhiezer46,Schlomann61,Sparks61}
Besides that three-particle processes may produce a spectacular quantum effect:
spontaneous magnon decay, which leads to a finite magnon lifetime
even at $T=0$. \cite{Zhitomirsky13} Theoretical predictions of spontaneous magnon
decay were made for dipolar ferromagnets, \cite{Lymar72,Syromyatnikov10,Chernyshev12}
for easy-plane ferromagnets, \cite{Baryakhtar79,Stephanovich11} and various
noncollinear antiferromagnets, see literature cited in Ref.~[\onlinecite{Zhitomirsky13}].
At the moment there are only a few experimental evidences of spontaneous magnon decay.
\cite{Stone06,Masuda10,Kurebayashi11,Oh13} Therefore, a natural question to ask
theoretically is what are the physical conditions that can enhance the magnon decay rate.
In the present work we focus on the role of low dimensionality in the magnon
decay and specifically consider whether the decay rate is enhanced in the quasi
one-dimensional (1D) geometry. This question was previously studied in the context
of quantum disordered magnets, \cite{Zhitomirsky06} but has not so far been
investigated for ordered magnetic systems. Our study is motivated, in part, by
a prominent example of quasi-1D easy-plane ferromagnet $\rm CsNiF_3$.
\cite{Steiner76} We investigate the feasibility of observation of spontaneous magnon
decays in inelastic neutron scattering experiments on this material.
The paper is organized as follows. In Sec.~II we formulate the spin model and
give necessary details of the $1/S$ spin-wave expansion. Sections~III and IV are devoted
to the discussion of the magnon damping in the quasi-1D and the 3D case, respectively.
Section~V considers the case of a weak antiferromagnetic coupling between ferromagnetic
chains and Sec.~VI gives our conclusions.

\section{Model}

\begin{figure}[tbp]
\centerline{
\includegraphics[width=30mm, height=45mm]{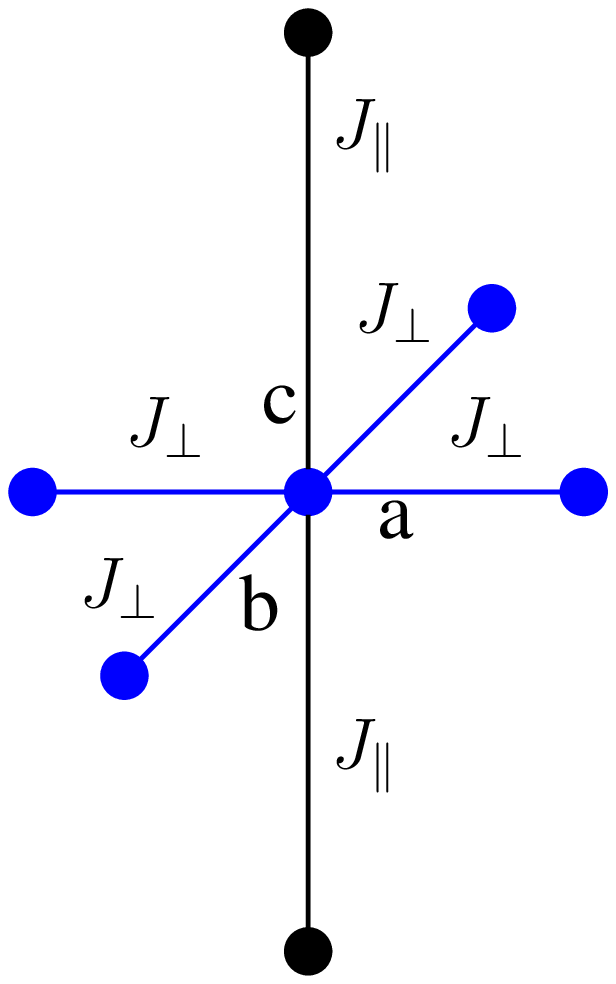}
\hspace*{1mm}
\includegraphics[width=35mm, height=50mm]{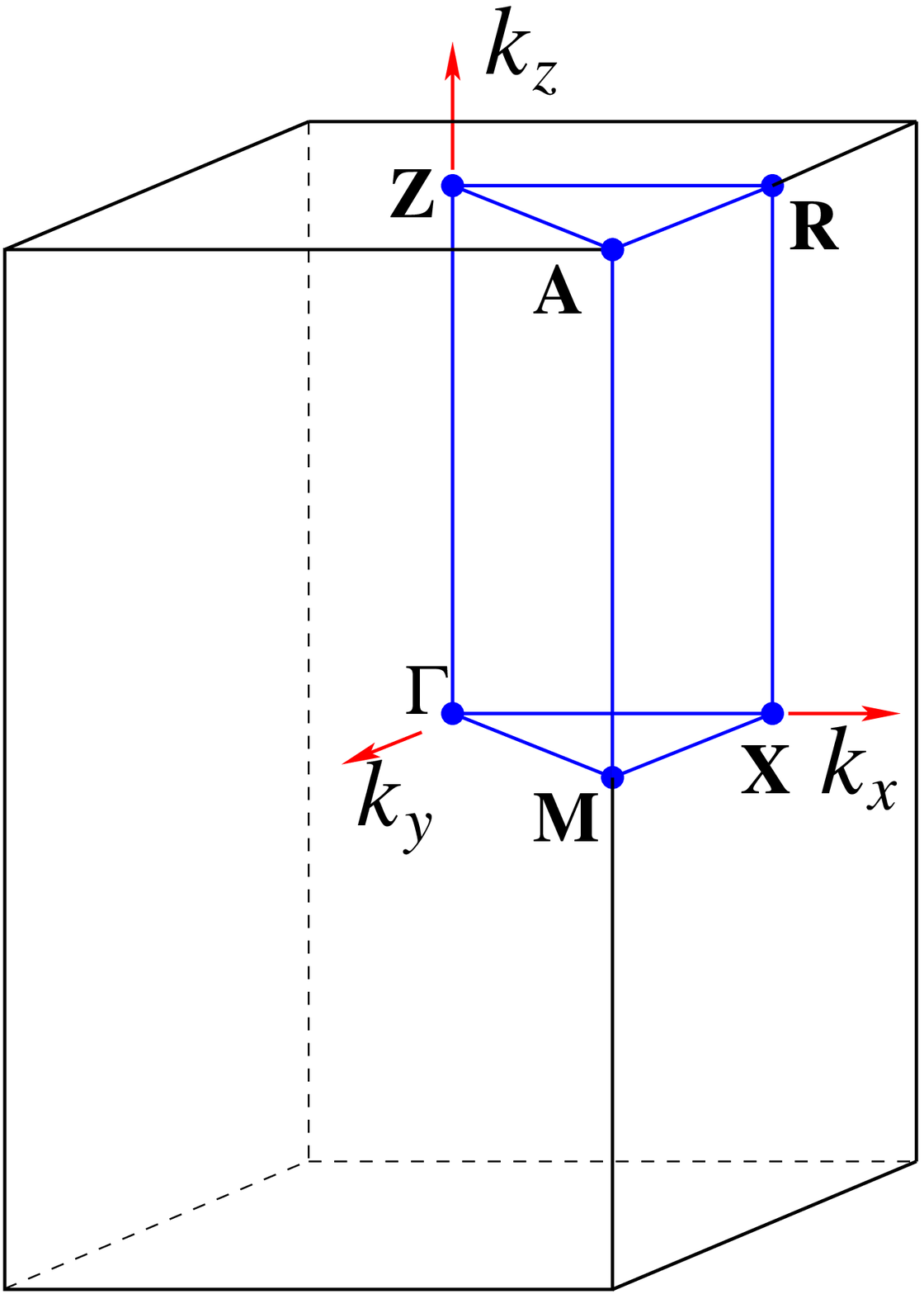}}
\caption{(Color online) Sketch of the unit cell of a tetragonal ferromagnet with nearest-neighbor
exchange interactions (left panel) and its Brillouin zone with notations for high
symmetry points (right panel).}
\label{fig1}
\end{figure}

We consider a Heisenberg ferromagnet with the easy-plane single-ion anisotropy
described by the spin Hamiltonian
\begin{eqnarray}
&&\hat{\mathcal H} = - J_\parallel\sum _i {\bf S}_i\cdot {\bf S}_{i+1} -
J_\perp \sum _{\langle ij\rangle} {\bf S}_i\cdot {\bf S}_j
\nonumber \\
&& \mbox{\qquad} + D\sum_i (S_i^z)^2  - H \sum_i S_i^z \ .
\label{HFM}
\end{eqnarray}
The nearest-neighbor exchange interactions consist of coupling $J_\parallel$
along chains parallel to the $z$-axis and inter-chain coupling  $J_\perp$.
Without loss of generality, we consider a square-type arrangement of chains in
the $x$--$y$ plane, see Fig.~\ref{fig1}. The choice $J_\parallel \simeq J_\perp$
corresponds to a 3D ferromagnet, $J_\parallel \gg J_\perp$---to a quasi-1D magnet,
whereas for  $J_\parallel \ll J_\perp$  a quasi-2D case is recovered. Quasi-1D
ferromagnetic material  CsNiF$_3$ has  a significant  easy-plane anisotropy
with $D\approx 0.32J_\parallel$.\cite{Steiner76} Motivated by this experimental
example  we fix in the following  $D\equiv 0.3J$, where $J$ is the largest of the
two exchange constants $J=\textrm{max}(J_\parallel, J_\perp)$. Since we are interested
in the behavior of high-energy magnons with $\varepsilon_{\bf k} \sim J_\parallel$
we are justified to neglect the much weaker dipolar interactions in the Hamiltonian
\eqref{HFM}.

We investigate the transverse field geometry with external field applied
along the hard axis and the ferromagnetic magnetization tilted
from the easy plane by angle $\theta$:
\begin{equation}
\sin\theta = \frac{H}{H_c}\,, \quad H_c = 2DS\,.  
\label{theta0}
\end{equation}
Above the critical field $H_c$ the ordered moments become completely aligned
with the hard axis. Note, that the critical field does not depend on
the ferromagnetic exchanges in (\ref{HFM}). As a result, the strength of the three-magnon
vertex stays remains unchanged under variation of $J_\perp/J_\parallel$, 
see Eq.~\eqref{decV} below, and the magnon damping at fixed $H$ solely 
depends on the magnon dispersion $\varepsilon_{\bf k}$ and its dimensionality.

To study excitations in the model \eqref{HFM}  we use the transformation from spins
to bosons introduced by Holstein and Primakoff. \cite{Holstein40} As usually,
the Holstein-Primakoff transformation is applied in the local frame such
that the local $z$-axis is collinear with a spin on a given site.
After performing a few standard steps \cite{Zhitomirsky13,Stephanovich11}
including expansion of square roots and
subsequent  Fourier and Bogolyubov transformations of  magnon operators
one obtains a spin-wave Hamiltonian structured in powers of $1/S$:
\begin{eqnarray}
\hat{\mathcal H}=\sum_{\bf k}\varepsilon_{\bf k} b_{\bf k}^\dagger b_{\bf k}
+\frac 12 \sum_{{\bf k},{\bf q}}V_{{\bf k},{\bf q}}
\Bigl[b_{\bf q}^\dagger b_{{\bf k}-{\bf q}}^\dagger b_{\bf k} + \textrm{h.c.}\Bigr]+...
\label{HH}
\end{eqnarray}
Here the magnon energy is $\varepsilon_{\bf k}=O(S)$,  the three-particle (cubic)
vertex responsible for spontaneous decays is $V_{{\bf k},{\bf q}} = O(S^{1/2})$ 
and ellipsis stand for the higher order terms.

An explicit expression for the harmonic magnon energy is
\begin{eqnarray}
\varepsilon_{\bf k} & = & 2S\sqrt{A_{\bf k}(A_{\bf k}\!+D\cos^2\!\theta)},
\ \ \gamma_{\bf k}= {\textstyle \frac{1}{2}} (\cos k_x\!+\cos k_y),
\nonumber \\
& & A_{\bf k} = J_\parallel(1-\cos k_z) + 2J_\perp(1-\gamma_{\bf k}).
\label{Ek}
\end{eqnarray}
The decay vertex is given by
\begin{equation}
V_{{\bf k},{\bf q}} = D\sqrt{\frac{S}{2}}\,\sin 2\theta
\bigl(g_{{\bf k},{\bf q},{\bf q'}}+f_{{\bf q},{\bf q'},{\bf k}}+f_{{\bf q'},{\bf q},{\bf k}}\bigr),
\label{decV}
\end{equation}
where ${\bf q'}=$ ${\bf k}$ - ${\bf q}$, $f_{1,2,3}=(u_1+v_1)(u_2u_3+v_2v_3)$,
$g_{1,2,3}=(u_1+v_1)(u_2v_3+v_2u_3)$, and $u_{\bf k}$, $v_{\bf k}$ are the Bogolyubov coefficients:
$$
u_{\bf k}^2 - v_{\bf k}^2=1 , \qquad 2u_{\bf k}v_{\bf k}=-DS\cos^2\theta/\varepsilon_{\bf k}.
$$
Note, that the vertex \eqref{decV} has a nonmonotonous  dependence on magnetic field:
$V_{{\bf k},{\bf q}} \propto H \sqrt{H_c^2-H^2}$, resulting in a strongest amplitude 
for magnon decay at $H/H_c = 1/\sqrt{2}$.

\begin{figure}[tbp]
\centerline{
\includegraphics[width=0.6\columnwidth]{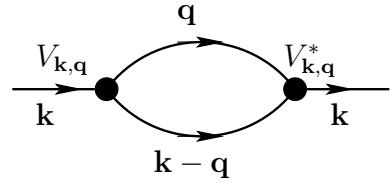}
}
\caption{(Color online) The self-energy diagram corresponding
to the considered two-magnon decay process.}
\label{dia}
\end{figure}

For a weakly interacting magnon gas, the magnon decay rate is given by the imaginary
part of the self-energy diagram shown in Fig.~\ref{dia} and coincides with 
the Fermi's golden-rule expression:
\begin{equation}
\Gamma_{\bf k}=\frac{\pi}{2} \sum_{\bf q}V^2_{{\bf k},{\bf q}}\,
\delta(\varepsilon_{\bf k}-\varepsilon_{\bf q}-\varepsilon_{{\bf k}-{\bf q}})\,.
\label{Gk}
\end{equation}

The two-magnon decay processes for easy-plane ferromagnets were
theoretically studied in Refs.~[\onlinecite{Baryakhtar79,Stephanovich11}].
Their appearance is determined by two conditions:\cite{Zhitomirsky13} (i) 
presence of the cubic vertex in the magnon Hamiltonian \eqref{HH}, which is 
a direct consequence of the fully broken spin-rotational symmetry for a state 
with tilted magnetization, and (ii) fulfillment of the energy conservation 
condition for the two-magnon decays:
\begin{equation}
\varepsilon_{\bf k}=\varepsilon_{\bf q}+\varepsilon_{{\bf k}-{\bf q}}, 
\label{cons}
\end{equation}
where the harmonic magnon energy \eqref{Ek} can be safely used due to 
smallness of quantum effects in ferromagnets.
The same kinematic conditions allow also three-magnon decays, which are present in
an easy-plane ferromagnet already in zero field. \cite{Stephanovich11}
The amplitude of these processes is, however, rather small 
as they correspond to higher-order $1/S$  terms of the spin-wave expansion
and we shall not consider them in the following.

Since the cubic vertex \eqref{decV} depends only on $D$ and $H$,
the effect of dimensionality on the magnon decay rate \eqref{Gk}
in anisotropic ferromagnet is present only
via the varying dispersion $\varepsilon_{\bf k}$. The two-dimensional case
was investigated in detail in  our previous work.\cite{Stephanovich11}
In the following sections we calculate the magnon decay rate \eqref{Gk}
for quasi-1D and 3D cases.

\section{%
Quasi one-dimensional case}

We begin the analysis of the magnon decay for $J_\parallel\gg J_\perp$
by treating analytically the case of long-wavelength magnons. In this limit the decay rate
can be calculated perturbatively because of smallness of interaction among low-energy excitations
and due to reduction of the phase-space volume available for decay processes.
Note that at small momenta $k,q\ll 1$, the decay vertex \eqref{decV} has the standard ``hydrodynamic''
form  $V_{{\bf k},{\bf q}} \propto \sqrt{kqq'}$. As a result, the
long-wavelength excitations exhibit a usual 3D asymptote
$\Gamma_{\bf k}$ $\propto k^5$
for the decay rate, \cite{Zhitomirsky13} because the dispersion $\varepsilon_{\bf k}$
is eventually three dimensional.
Therefore, the proper question to be addressed analytically
is how the coefficient in the $k^5$-law depends on a small parameter
$J_\perp/J_\parallel$.

An analytical derivation  of the low-energy asymptote for $\Gamma_{\bf k}$ closely follows
a similar computation for 2D or 3D magnetic system with three-particle vertices.\cite{Zhitomirsky13}
Below we present only the essential steps. Expanding \eqref{Ek} in small $k$ one obtains
\begin{eqnarray}
&& \varepsilon_{\bf k} \approx c\sqrt{k_z^2+jk_\perp^2}\left[1 + \alpha k_z^2\right]\,,
\label{Eklow} \\
& & c=S\cos\theta\sqrt{2DJ_\parallel}\,, \quad \alpha = J_\parallel/(4D\cos^2\theta)-1/24\,,
\nonumber
\end{eqnarray}
where  $j = J_\perp/J_\parallel$ and
$k_\perp^2=k_x^2+k_y^2$. Strictly speaking, the above expression for $\alpha$
looses its validity for $k_z\ll k_\perp$. However, as we shall see shortly, the region of
interest in the quasi-1D case is  $k_\perp\ll k_z$, which justifies
Eq.~\eqref{Eklow}.

Selecting the momentum of an incident magnon on the $z$-axis, ${\bf k} = (0,0,k)$,
we obtain in the same approximation
\begin{equation}
\label{Econserv}
\varepsilon_{\bf k}-\varepsilon_{\bf q}-\varepsilon_{{\bf k}-{\bf q}}\approx -
\frac{cjk}{2q_z(k-q_z)}\left(q_\perp^2-q_0^2\right),
\end{equation}
where $q_0^2=6\alpha q_z^2(k-q_z)^2/j$.
Substituting  \eqref{Econserv} into the expression for the decay rate \eqref{Gk}
and performing separate integration of $q_\perp$ and $q_z$ we obtain
the following long-wavelength asymptote in the quasi-1D case
\begin{equation}
\label{Gk1}
\Gamma_{\bf k}=a\frac{J_\parallel^2}{J_\perp}\tan^2\theta \ k^5\,,
\end{equation}
where a dimensionless constant  is $a\sim 10^{-3}$ and $\theta$ is the canting angle
\eqref{theta0}. In a 3D case for $J_\parallel$=$J_\perp$ a similar computation yields
\cite{Baryakhtar79}
\begin{equation}
\label{Gk3}
\Gamma_{\bf k}=\frac{3J_\parallel}{160\pi}\tan^2\theta \ k^5\,.
\end{equation}
Thus, in the quasi-1D case the damping of acoustic magnons is inversely proportional
to a small $J_\perp$ and is, therefore, parametrically enhanced compared to damping of
acoustic magnons in 3D. We have verified such an enhancement by a direct numerical integration
of Eq.~\eqref{Gk}.

\begin{figure}[t]
\centering
\includegraphics[width=0.9\columnwidth]{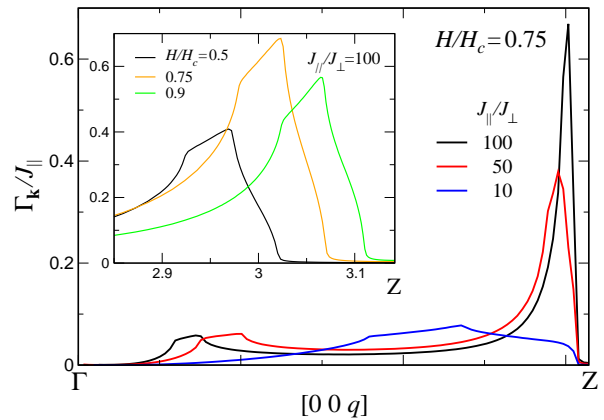}
\caption{(Color online) Decay rate for magnons in a quasi-1D ferromagnet for
momenta along the $\Gamma$Z direction for different $J_\parallel/J_\perp$ (legend)
and $H/H_c=$0.75. Inset shows
the field dependence of the peak in $\Gamma_{\bf k}$ near the Z-point.}
\label{figg2}
\end{figure}

Figure \ref{figg2} shows the magnon decay rate \eqref{Gk} evaluated numerically
at representative field value $H/H_c=$0.75  in the $\Gamma$Z direction. The curves
correspond to three $J_\parallel/J_\perp$ ratios portraying crossover from
a strong $J_\parallel/J_\perp=100$ to a weak $J_\parallel/J_\perp=10$ quasi
one-dimensionality.
The two curves corresponding to $J_\parallel/J_\perp=50,100$ exhibit
large peaks in $\Gamma_{\bf k}$ near the Brillouin zone boundary, which originate
from a 1D Van Hove singularity in the spectrum. A more detailed structure of this peak
on the inset illustrates the role of 3D coupling, which cuts off the square-root divergence
of the peaks and restores a 3D Van Hove singularities at the boundary of the decay region
and for the saddle points in the continuum.
Similar 3D Van Hove singularities are also prominent for small momenta
towards the $\Gamma$ point.
Importantly, the height of the peak in $\Gamma_{\bf k}$ near the $Z$-point decreases rapidly
as the magnon dispersion becomes more and more 3D.
This demonstrates that the decays along
$\Gamma$Z direction in the Brillouin zone are most prominent for the quasi-1D case.

The magnetic field dependence of the magnon damping in the region, where
$\Gamma_{\bf k}$ is largest, is illustrated on the inset of Fig.~\ref{figg2}.
One can see a nonmonotonous field dependence of the peak height:
$\Gamma_{\bf k}$ is smallest at $H/H_c=0.5$, it is largest at $H/H_c=0.75$,
while it again goes down at $H/H_c= 0.9$.
Such a behavior is related to the field dependence of the decay vertex \eqref{decV},
which is zero at $H=0$  and $H=H_c$ and has a maximum at $H/H_c=1/\sqrt{2}\approx 0.707$.

\section{Three-dimensional case}

\begin{figure}[b]
\centering
\includegraphics[width=0.9\columnwidth]{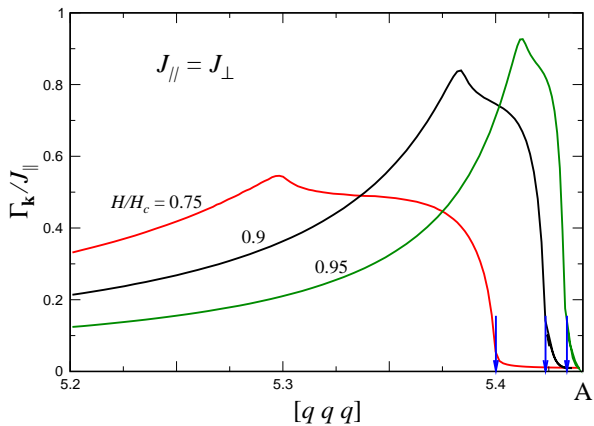}
\caption{(Color online) Decay rate for magnons in a 3D ferromagnet ($J_\perp/J_\parallel=1$)
with momenta along the $\Gamma$A direction for  different magnetic fields
(figures near curves). Vertical arrows mark the magnon decay boundaries
${\bf k}^*$, obtained from the kinematic condition
$\varepsilon_{\bf k} = 2\varepsilon_{{\bf k}/2}$.}
\label{figg3}
\end{figure}

Spontaneous magnon decays for a 2D easy-plane ferromagnet in transverse magnetic
field were studied in our previous work. \cite{Stephanovich11} In particular,
the decay rate $\Gamma_{\bf k}$ exhibits logarithmic peaks, which are determined
by saddle-point Van Hove singularities in the two-magnon density of states.
Changing the type of anisotropy, exchange versus single-ion, has no significant effect on
the decay dynamics. Taking into account 1D (Sec.~III) and 2D results \cite{Stephanovich11}
one can speculate that the magnon decay in 3D shows no major enhancement
and, thus, should be small compared to low dimensional magnets.

To study  the magnon damping in 3D we fix the exchange ratio to $J_\parallel/J_\perp=1$
as an example.
It is instructive to  consider in this case magnons with momenta belonging to the cubic
diagonal, the $\Gamma A$ direction. Numerical results for $\Gamma_{\bf k}$ along this axis
look very similar to  Fig.~\ref{figg2} including a surprisingly high peak near the A point.
A zoom into this region is shown in Fig.~\ref{figg3} for different values of an applied
magnetic field. The overall shape  of numerical data is qualitatively similar to the results
shown on the inset of Fig.~\ref{figg2} for the quasi-1D case. The peak in $\Gamma_{\bf k}$
is most prominent for $H/H_c > 0.75$, while at smaller fields it is much less pronounced.
The arrows show the magnon decay boundaries, obtained from the kinematic condition
$\varepsilon_{\bf k}=2\varepsilon_{{\bf k}/2}$, see further details
in  Ref.~[\onlinecite{Zhitomirsky13}].
The above fact actually means that the predominant decay channel for a magnon  in the vicinity
of the damping peak is a decay into two magnons  with equal momenta lying on the same cubic diagonal.
With increasing $H$ the decay region extends further towards the $A$ point, ${\bf k}_0=(\pi,\pi,\pi)$,
the two magnons emitted in a decay process become
close to ${\bf k}_0/2$. A remarkable property of the nearest-neighbor
magnon dispersion (\ref{Ek}) near ${\bf k}_0/2$
is that $\varepsilon_{\bf k}$ is almost perfectly flat with the exception of a few special directions.
Such an effective dimensionality reduction is responsible for the enhanced
two-magnon density of states (DOS), which in turn leads to large values of the magnon decay rate \eqref{Gk}.

\begin{figure}[t]
\centering
\includegraphics[width=0.9\columnwidth]{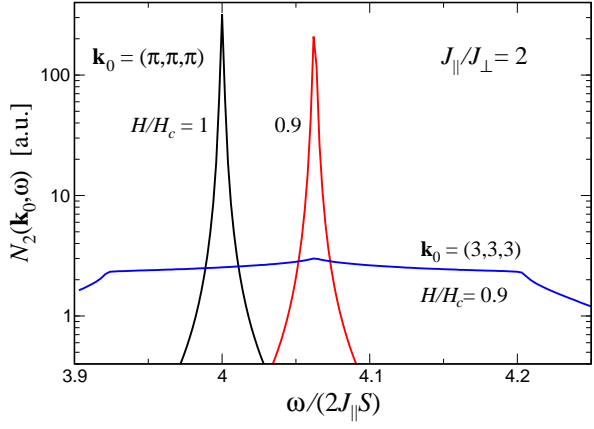}
\caption{(Color online) The two-magnon density of states for different values of magnetic field (figures near curves) and $J_\parallel/J_\perp=2$. Curve for $H/H_c=0.9$ and ${\bf k}_0=(3,3,3)$ is shown for comparison.}
\label{figg4}
\end{figure}

To check the above scenario for the magnon damping enhancement in 3D case
we calculate the two-magnon DOS:
\begin{equation}
\label{dos}
N_2({\bf k},\omega)=\sum_{\bf q}\delta(\omega-\varepsilon_{{\bf k}/2+{\bf q}}-\varepsilon_{{\bf k}/2-{\bf q}})\,.
\end{equation}
At $H=H_c$ the DOS exhibits a delta-peak for ${\bf k} = {\bf k}_0$. Indeed, in that case the magnon energy
(\ref{Ek}) is given by a sum of the cosine harmonics. For ${\bf k}={\bf k}_0$ one has $\cos(k_0/2+q)\to\sin q$ and
the sum of  two magnon energies on the r.h.s. of \eqref{dos} yields a constant. Hence,
\begin{equation}
\label{dos1}
N_2({\bf k}_0, \omega)=\delta(\omega-\varepsilon_{{\bf k}_0})\,, \quad \varepsilon_{{\bf k}_0}=2J_0S\,,
\end{equation}
where $J_0=2J_\parallel+4J_\perp$. For $H\alt H_c$, we expand $E({\bf k}_0,{\bf q}) =
\varepsilon_{{\bf k}_0/2+{\bf q}}-\varepsilon_{{\bf k}_0/2-{\bf q}}$ in small
$q$ as follows
\begin{eqnarray}
&& E({\bf k}_0,{\bf q})\approx 2S\sqrt{J_0J_D} \left[1 - a_{\bf q}^2\, \frac{(J_D-J_0)^2}{2J_D^2J_0^2} \right] ,
\label{dos2}  \\
&& J_D= J_0 + 2D\cos^2\theta\,, \quad a_{\bf q} = J_\parallel q_z+ J_\perp(q_x+q_y)\,.
\nonumber
\end{eqnarray}
The dependence of $E({\bf k}_0,{\bf q})$ on ${\bf q}$ enters only via a linear combination $a_{\bf q}$ \eqref{dos2}.
Thus, neglecting higher-order terms one finds an effective 1D dispersion of the decay surface
$E({\bf k}_0,{\bf q})= \omega$.
As a result, an integration of the delta-function in Eq.~\eqref{dos} generates a conventional 1D square-root
Van Hove singularity in the DOS:
\begin{equation}
\label{dos3}
N_2({\bf k}_0, \omega)\simeq \frac{1}{\cos\theta\sqrt{|\omega-2\varepsilon_{{\bf k}_0/2}|}}.
\end{equation}
At $H=H_c$ ($\theta=\pi/2$) the square-root peak transforms into the delta-peak discussed above.
Note also, that $\varepsilon_{{\bf k}_0}<2\varepsilon_{{\bf k}_0/2}$ for $H<H_c$ with the equality
(signifying a fulfilment of the kinematic decay condition) reached only at $H=H_c$.
The above analytical results for DOS can be compared with the direct
numerical evaluation of \eqref{dos} presented in Fig.~\ref{figg4}.

For small departures from ${\bf k}_0$ point, $H<H_c$, the two-magnon energy
$E({\bf k},{\bf q})$  acquires full dispersion on components of $\bf q$ and
the Van Hove singularity \eqref{dos3} is smeared. Still, an enhanced DOS at the former
peak position survives for a range of values of $\bf k$. As an illustration
Figure~\ref{figg4} shows the numerical result for ${\bf k}=(3,3,3)$ at $H=0.9H_c$.
A remnant peak  in DOS (mind the logarithmic scale in Fig.~\ref{figg4})
leads to a stronger decay rate $\Gamma_{\bf k}$ once magnetic
field approaches $H_c$ and magnons with momenta close to ${\bf k}_0$
become unstable.

\section{Antiferromagnetic interchain coupling}

As was mentioned in  the Introduction, our study is in a large part motivated by
the quasi-1D  ferromagnet $\rm CsNiF_3$. \cite{Steiner76} Magnetic Ni$^{2+}$ ions
($S=1$) are arranged in this material in a hexagonal lattice with a ferromagnetic exchange
$J_\parallel \approx 24$~K along the $c$-axis and  an antiferromagnetic interchain
coupling $|J_\perp/J_\parallel|\sim 10^{-2}$.  The strength of the single-ion
anisotropy in $\rm CsNiF_3$ is $D \approx 8$~K.
The  antiferromagnetic transition in $\rm CsNiF_3$ takes place at $T_N=2.5$~K,
however, magnetic moments on adjacent chains do not form the 120$^\circ$  structure
with the propagation wave-vector $(1/3,1/3,0)$ expected for a triangular geometry, but rather
order collinearly with ${\bf Q}=(1/2,0,0)$. \cite{Steiner74}
The collinear order has been explained by  a competition between the antiferromagnetic
exchange and the long-range dipolar interactions. \cite{Baehr96}
In this Section we consider the effect of the sign on an interchain coupling on the
magnon decay in quasi-1D chains.
We shall use a simplified model \eqref{HFM} with $J_\perp<0$ assuming a tetragonal arrangement of
chains and neglecting dipolar interactions. In this model, spin chains are still ordered
ferromagnetically while an ordering  between chains is described by the wavevector
${\bf Q}=$ $(\pi,\pi,0)$ and corresponds to a two sublattice  antiferromagnetic structure.

\begin{figure}[t]
\centering
\includegraphics[width = 0.99\columnwidth]{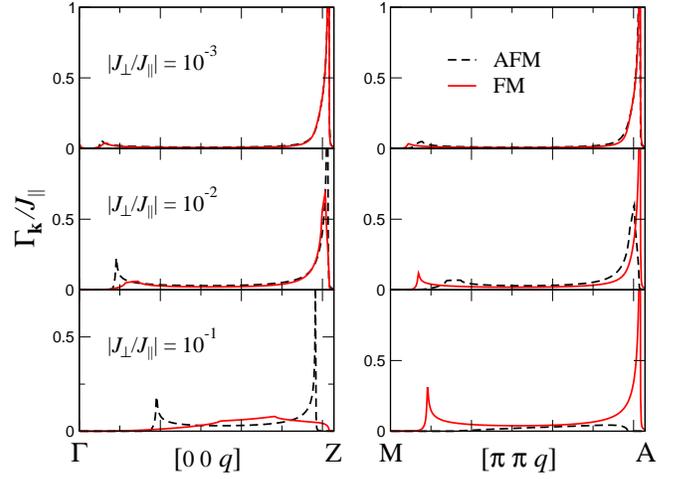}
\caption{(Color online) Comparison of magnon decay rates for ferromagnetic (FM, $J_\perp>0$, solid line)
and antiferromagnetic (AFM, $J_\perp<0$, dashed line) for $H/H_c=0.8$.
Left column - $\Gamma$Z direction, right column - MA direction.
The ratios $|J_\perp/J_\parallel |$ are 10$^{-3}$ (upper panels), 10$^{-2}$ (middle panels),
and 10$^{-1}$ (lower panels).}
\label{figg5}
\end{figure}

We again assume that an external magnetic field is oriented
along the hard axis. Theoretical calculations in this case become very similar to
the spin-wave theory for a Heisenberg square-lattice antiferromagnet.
\cite{Mourigal10,Zhitomirsky99}
The two magnetic sublattices tilt from the easy plane by an angle $\theta$
\begin{equation}
\sin\theta = \frac{H}{2S(D+4|J_\perp|)} \,.
\label{theta1}
\end{equation}
In contrast to the ferromagnetic case \eqref{theta0}, the critical field $H_c=2S(D+4|J_\perp|)$
depends on an antiferromagnetic exchange $J_\perp$.
After performing a standard spin-wave calculation, see Sec.~II,  we obtain the magnon energy
\begin{eqnarray}
\varepsilon_{\bf k} & = &
2S\sqrt{A_{\bf k}(A_{\bf k}+D\cos^2\theta-4|J_\perp|\gamma_{\bf k}\cos^2\theta)}\,,
\nonumber \\
& & A_{\bf k} = J_\parallel(1-\cos k_z) + 2|J_\perp|(1+\gamma_{\bf k})\,.
\label{Ekafm}
\end{eqnarray}
Due to a staggered canting of spins in the antiferromagnetic structure the three-particle
term in (\ref{HH}) contains now  ${\bf q}'= {\bf k} - {\bf q} + {\bf Q}$
instead of ${\bf k} - {\bf q}$. The explicit expression for the
decay vertex is
\begin{equation}
V_{{\bf k},{\bf q}} = \sqrt{2S}\,\sin\theta\cos\theta
\bigl(g_{{\bf k},{\bf q},{\bf q'}}+f_{{\bf q},{\bf q'},{\bf k}}+f_{{\bf q'},
{\bf q},{\bf k}}\bigr),
\label{decV1}
\end{equation}
where $f_{1,2,3}=\lambda_{1,2,3}(u_1+v_1)(u_2u_3+v_2v_3)$,
$g_{1,2,3}=\lambda_{1,2,3}(u_1+v_1)(u_2v_3+v_2u_3)$, and $\lambda_{\bf k}=$ $D-4J_\perp \gamma_{\bf k}$.

The magnon decay rates for ferromagnetic ($J_\perp>0$) and antiferromagnetic ($J_\perp<0$) sign
of the interchain coupling are compared in Fig.~\ref{figg5}.
The presented results illustrate a crossover from an extreme $|J_\perp/J_\parallel|=0.001$
to a moderate $|J_\perp/J_\parallel|=0.1$ quasi one-dimensionality.
Plots in the left column show $\Gamma_{\bf k}$ for momenta on the $\Gamma$Z-line,
while  the right column corresponds to the MA cut, which includes the
antiferromagnetic vector ${\bf Q}=$ $(\pi,\pi,0)$, see Fig.~\ref{fig1} for the notations.
For a very weak interchain coupling $|J_\perp/J_\parallel|=10^{-3}$, there is no
significant difference in the magnon damping $\Gamma_{\bf k}$ between two signs
of $J_\perp$ and also between two lines. Overall, $\Gamma_{\bf k}$ exhibits
the same behavior as results in Fig.~\ref{figg2}
calculated for different values of magnetic field.  In particular,
high peaks are present near the BZ boundary for both lines, $\Gamma$Z and MA.

Some differences start to develop for $|J_\perp/J_\parallel|\sim 0.01$  and become quite significant at
$|J_\perp/J_\parallel|\sim 0.1$. The magnon dispersion between chains becomes more substantial and
plays a more prominent role in the energy conservation. As a result,  for the ferromagnetic interchain coupling
$J_\perp>0$, a stronger magnon damping with two characteristic  peaks is present for magnons on the MA-line,
whereas $\Gamma_{\bf k}$ on the $\Gamma$Z-line is significantly smaller for $J_\perp/J_\parallel = 0.1$.
For the antiferromagnetic interchain coupling $J_\perp<0$, one can observe an opposite tendency.
In fact, there is a remarkable mirror symmetry between plots on the left and on the right with simultaneous
sign change of $J_\perp$. It is related to the fact that the position of the acoustic magnon branch
alters its place between $\Gamma$Z- and MA-line with the sign reversal. Overall, most favorable conditions
for observing spontaneous magnon decay,
{\it i.e.}, large $\Gamma_{\bf k}$ for extended region in the momentum space, are found for
$|J_\perp/J_\parallel|=0.01$, a value close to exchange ratio in $\rm CsNiF_3$.

\section{Conclusions}

In summary, magnetic excitations in an easy-plane ferromagnet placed in
a transverse magnetic field become intrinsically damped at $T=0$ due to
two-magnon decays. \cite{Baryakhtar79,Stephanovich11}
We have studied the effect of dimensionality on the magnon decay rate
for such an ordered magnetic system.
For weak interchain coupling the decay rate $\Gamma_{\bf k}$ is strongest
in the vicinity of the Brillouin zone boundary exhibiting a peak
$\Gamma_{\bf k} \sim 0.4$--$0.7J_\parallel$. Such a peak is related
to the 1D-like Van Hove singularity in the two-magnon density of states
and its height needs to be compared to the
characteristic energy of magnons at the Brillouin zone boundary
$\varepsilon_{\bf k} = 4J_\parallel S$. For $S = 1/2$ and $S = 1$
the decay rate is a sizeable fraction of the magnon energy.
Therefore, spontaneous magnon decays can be observed
in the neutron-scattering experiments as a significant line broadening of the
zone boundary magnons. In particular, our results for a model system \eqref{HFM}
with ferromagnetic spin chains, which are weakly coupled antiferromagnetically,
indicate that the spontaneous magnon decay should be prominent in
the quasi 1D easy-plane ferromagnet $\rm CsNiF_3$.  \cite{Steiner76}

Somewhat surprisingly, we have found a large decay rate $\Gamma_{\bf k}$ also in a 3D case
$J_\perp \sim J_\parallel$ for certain magnon momenta $\bf k$.
The increase in $\Gamma_{\bf k}$ is again rooted in the two-magnon density of states
$N_2({\bf k},\omega)$, see Eq.~\eqref{dos}, which develops a peak due to a very weak ${\bf q}$-dependence of
the two magnon energy $\varepsilon_{{\bf k}/2+{\bf q}} + \varepsilon_{{\bf k}/2-{\bf q}}$.
The latter property is a consequence of the exchange coupling only between the
nearest neighbors. Being a model assumption this property is nevertheless satisfied with
good accuracy in many magnetic insulators.

Overall, we conclude that lower dimensionality has a pronounced  effect on spontaneous magnon decay
by means of an enhanced two-magnon DOS. In the quasi-1D limit, the Van Hove singularities in the DOS
become largest and correspond to a strong damping of magnons in the vicinity of the decay threshold
boundary. Thus, small shift in the $\bf k$-space may lead to striking changes in the behavior
of the corresponding magnon modes. Our theoretical results call for inelastic neutron-scattering
measurements of magnon lifetime in spin-chain materials. Materials with ferromagnetic chains
are, especially, suitable for such experiments since the required magnetic fields $H\sim \max\{D,J_\perp\}$
can be rather small and the effect of magnon decay can be clearly distinguished from the
spinon physics, which is present for antiferromagnetic chains.


\begin{thebibliography}{99}

\bibitem{Bloch30}
F. Bloch, Z. Physik \textbf{61}, 206 (1930).

\bibitem{Holstein40}
T. Holstein and H. Primakoff, Phys. Rev. \textbf{58}, 1098 (1940).

\bibitem{Oguchi60}
T. Oguchi, Phys. Rev. \textbf{117}, 117 (1960).

\bibitem{Akhiezer46}
A. I. Akhiezer, J. Phys. (U.S.S.R.)  \textbf{10}, 217 (1946).

\bibitem{Anderson}
P. W. Anderson, \textit{Basic notions of condensed matter
physics} (Benjamin-Cummings, Menlo Park, 1984).

\bibitem{Ohyama93}
T. Ohyama and H. Shiba,
J. Phys. Soc. Jpn. \textbf{62}, 3277 (1993).

\bibitem{Zhitomirsky99}
M. E. Zhitomirsky and A. L. Chernyshev,
Phys. Rev. Lett. {\bf 82}, 4536 (1999).


\bibitem{Harris71}
A. B. Harris, D. Kumar, B. I. Halperin, and P. C. Hohenberg,
Phys. Rev. B \textbf{3}, 961 (1971).

\bibitem{Schlomann61}
E. Schl\"omann, Phys. Rev. \textbf{121}, 1312 (1961).

\bibitem{Sparks61}
M. Sparks, R. Loudon, and C. Kittel,
Phys. Rev. \textbf{122}, 791 (1961).

\bibitem{Zhitomirsky13}
M. E. Zhitomirsky and A. L. Chernyshev,
Rev. Mod. Phys. \textbf{85}, 219 (2013).

\bibitem{Lymar72}
V. I. Lymar and Yu. G. Rudoi,
Theor. Math. Phys. \textbf{11}, 376 (1972).

\bibitem{Syromyatnikov10}
A. V. Syromyatnikov,
Phys. Rev. B \textbf{82}, 024432 (2010).

\bibitem{Chernyshev12}
A. L. Chernyshev,
Phys. Rev. B \textbf{86}, 060401(R) (2012).

\bibitem{Baryakhtar79}
V. G. Baryakhtar, A. I. Zhukov, and D. A. Yablonskii,
Fiz. Tverd. Tela \textbf{21}, 776 (1979)
[Sov. Phys. Solid State \textbf{21}, 454 (1979)].

\bibitem{Stephanovich11}
V. A. Stephanovich and M. E. Zhitomirsky,
Europhys. Lett., {\bf 95}, 17007 (2011).

\bibitem{Stone06}
M. B. Stone,  I. A. Zaliznyak, T. Hong, C. L. Broholm, and
D. H. Reich,  Nature \textbf{440}, 187 (2006).

\bibitem{Masuda10}
T. Masuda, S. Kitaoka, S. Takamizawa, N. Metoki, K.
Kaneko, K. C. Rule, K. Kiefer, H. Manaka, and H. Nojiri,
Phys. Rev. B \textbf{81}, 100402(R) (2010).

\bibitem{Kurebayashi11}
H. Kurebayashi, O. Dzyapko, V. E. Demidov, D. Fang,
A. J. Ferguson, and S. O. Demokritov,
Nat. Mater. \textbf{10}, 660 (2011).

\bibitem{Oh13}
J. Oh, M. D. Le, J. Jeong, J. Lee, H. Woo, W.-Y. Song, T. G. Perring,
W. J. L. Buyers, S.-W. Cheong, and J.-G. Park,
Phys. Rev. Lett. \textbf{111}, 257202 (2013).

\bibitem{Zhitomirsky06}
M. E. Zhitomirsky,  Phys. Rev. B \textbf{73}, 100404(R)
(2006).

\bibitem{Steiner76}
M. Steiner, J. Villain, and C. G. Windsor,
Adv. Phys. {\bf 25}, 87 (1976).

\bibitem{Steiner74}
M. Steiner, and H. Dachs, Solid State Commun. {\bf 14}, 841 (1974).

\bibitem{Baehr96}
M. Baehr, M. Winkelmann, P. Vorderwisch, M. Steiner, C. Pich,
and F. Schwabl, Phys. Rev. B {\bf 54}, 12932 (1996).

\bibitem{Mourigal10}
M. Mourigal, M. E. Zhitomirsky and A. L. Chernychev
Phys. Rev. B {\bf 82}, 144402 (2010).

\end{thebibliography}
\end{document}